# The Price and Cost of Bitcoin

John E. Marthinsen and Steven R. Gordon

## Abstract

Explaining changes in bitcoin's price and predicting its future have been the foci of many research studies. In contrast, far less attention has been paid to the relationship between bitcoin's mining costs and its price. One popular notion is the cost of bitcoin creation provides a support level below which this cryptocurrency's price should never fall because if it did, mining would become unprofitable and threaten the maintenance of bitcoin's public ledger. Other research has used mining costs to explain or forecast bitcoin's price movements. Competing econometric analyses have debunked this idea, showing that changes in mining costs follow changes in bitcoin's price rather than preceding them, but the reason for this behavior remains unexplained in these analyses. This research aims to employ economic theory to explain why econometric studies have failed to predict bitcoin prices and why mining costs follow movements in bitcoin prices rather than precede them. We do so by explaining the chain of causality connecting a bitcoin's price to its mining costs.

## Introduction

A significant body of research has been devoted to explaining changes in bitcoin's price and predicting its future.[1] Although this research has achieved a modicum of success in identifying the relatively stable price movements, none of them has explained bitcoin's significant price volatility.[2] Recent research has achieved more success using autoregressive statistical models, wavelet theory, and neural networks to explain bitcoin's price movement, at least in the short term, as based on past trends and investor sentiment. The feedback cycles and mechanisms these models capture are well suited to explaining bitcoin's price volatility but fall short in predicting bitcoin's longer-term behavior as they fail to incorporate any underlying causal mechanisms.

---

[1] See Tables 1, 2, and 3.

[2] Some research (for example, Estrada, 2017, Katsiampa, 2017, and Aalborg et al., 2019) has succeeded in modeling the degree of volatility in short and medium terms, but these models are mainly autoregressive and fail to capture or explain how bitcoin's price increased nearly eight-fold from mid-January 2017 to mid- December 2017, fell more than 80% from then until mid-December 2018, or increased more than 11-fold from mid-March 2020 to mid-March 2021; nor are they able to explain what caused the reversal in these outsized upward or downward trends.

In contrast to bitcoin price research, far less attention has been paid to the relationship between bitcoin's mining costs and its price. One popular notion, largely unsubstantiated and unchallenged, is that the cost of bitcoin creation provides a support level below which this cryptocurrency's price should never fall because if it did, mining would become unprofitable and threaten the maintenance of bitcoin's public ledger (Garcia et al., 2014). A related belief is that bitcoin's price must rise with an increase in the cost of production (e.g., mining expenses). Other research has used mining costs to explain or forecast bitcoin's price (Meynkhard, 2019; Hayes, 2019). Kristofek (2020) and Fantazzini & Kolodin (2020) employ econometric analyses to debunk this theory, showing that changes in mining costs follow changes in bitcoin's price rather than preceding them, but the reason for this behavior remains unexplained in these analyses.

This research aims to employ economic theory to explain why econometric studies have failed to predict bitcoin prices and why mining costs follow movements in bitcoin prices rather than precede them. We start with a review of prior research into the causes of bitcoin prices changes, followed by an economic explanation for its unpredictability. Finally, we explain the chain of causality connecting a bitcoin's price to its mining costs.

## Review of the Literature

### Economic Factors and the Bitcoin's Price

Foreign exchange pricing analyses often employ models based on the quantity theory of money (QTM) or purchasing power parity (PPP). These models are of little use with bitcoin because, currently, bitcoin is not a widespread unit of account or medium of exchange (Baur et al., 2018). Although some companies accept bitcoins, far fewer price their products and services in this cryptocurrency. Instead, they use the fiat currencies of their economic regions, such as dollars or euros, to price their sales items. If they accept bitcoin, customers convert these prices according to the spot exchange rate (i.e., bitcoin's price). As a result, it is impossible to create a price index for bitcoin similar to the consumer price index or implicit price index that exists for dollar-denominated products consumed or produced in the United States or for goods and services denominated in the local currency of any other country. The QTM and PPP models rely on such indices or substitutes for them, but none appear to be reliable for bitcoin.

Nevertheless, there are valid reasons to expect bitcoin's price to react to changes in economic and financial conditions. For example, increases in a nation's inflation rate relative to other countries or hyperinflation in any nation, all else being equal, should increase the demand for alternatives, such as bitcoin, and appreciate its price relative to the inflation-prone currencies. Table 1 identifies some of the most important studies addressing the impact of economic and financial variables on bitcoin's price. Except for a few tested variables, their results have not been robust.

| Table 1 | |
|---|---|
| **Economic and Financial Sources of Bitcoin Price Changes: US$ per Bitcoin** | |
| **Independent Variable** | **Results** |
| **Security Prices** | |
| Shanghai Stock Market Index | Positive (Bouoiyour & Selmi, 2015; Kristoufek, 2013; Kristoufek, 2015) Insignificant (Bouoiyour et al., 2015) |
| Dow Jones Index | Insignificant (Kristoufek, 2013) Positive* (Ciaian et al., 2016) |
| Financial Stress Index (FSI) | Insignificant except during Cypriot Crisis (Kristoufek, 2015) |
| S&P500, S&P600 (U.S. equity) | Insignificant (Baur et al., 2018) |
| Various Bloomberg U.S. corporate and government bond indices | Insignificant (Baur et al., 2018) |
| World stock market indices | Negative (Goczek & Skliarov, 2019) |
| **Currency Prices** | |
| Euro-U.S. dollar exchange rate (€/$) | Insignificant (Baur et al., 2018) Positive* (Ciaian et al., 2016) |
| Yen- U. S. dollar exchange rate (¥/$) | Insignificant (Baur et al., 2018) |
| Exchange rate of various currencies vs. U.S. dollar | Insignificant (Baur et al., 2018) |
| Trade-weighted dollar index | Insignificant (Baur et al., 2018) |
| Exchange rate volatility/safe haven | Insignificant (Baur et al., 2018) |
| Price of other cryptocurrencies | Positive (Nguyen et al., 2018) |
| **Commodity Prices** | |
| Oil prices | Insignificant (Ciaian et al., 2016; Kristoufek, 2013; Kristoufek, 2015; Baur et al., 2018) Positive* (Ciaian et al., 2016) |
| Natural gas | Insignificant (Baur et al., 2018) |
| Gold price | Insignificant (Baur et al, 2018; Bouoiyour & Selmi, 2015; Kristoufek, 2013; Kristoufek, 2015) |
| Silver | Insignificant (Baur et al., 2018) |
| * means significant in the short run only. | |

## Ecosystem and Algorithmic Factors Affecting Bitcoin's Price

Prior research has addressed the impact that elements of the bitcoin ecosystem have had on this cryptocurrency's price. Most of them, such as the number of transactions per period and number of addresses, act as proxies for the breadth and depth of the bitcoin market. As more individuals and business leaders become aware of and comfortable with the existence and potential worth of a cryptocurrency, such as bitcoin, the demand increases, causing its exchange rate value to appreciate.

Other ecosystem research seeks to determine how the mining ecosystem affects bitcoin's price. Bitcoin's hash rate, the total number of coins in circulation, the cost of mining, and the rate at which miners create new coins are all potential factors. Bitcoin's algorithm adjusts the difficulty of mining so that miners create new coins at a constant rate for approximately four years. Every 210,000 blocks, the algorithm reduces by half the rate of new coin issuance, an event that some researchers have perceived to be a mining-cost shock that should reduce the flow of bitcoin supplied to the currency markets. Table 2 identifies several studies that have explored how Bitcoin's ecosystem and algorithm have affected its price.

| Table 2 | |
|---|---|
| **Ecosystem and Algorithmic Sources of Bitcoin Price Changes: US$ per Bitcoin** | |
| **Independent Variable** | **Results** |
| Number of bitcoin transactions | Positive (Kristoufek, 2015; Ciaian et al., 2016) |
| Number of bitcoin addresses | Positive (Ciaian et al., 2016; Hau et al., 2021) |
| Estimate output volume | Negative (Bouoiyour et al., 2015; Bouoiyour & Selmi, 2015; Kristoufek, 2013; Kristoufek, 2015) |
| Trade-exchange ratio | Insignificant (Bouoiyour et al., 2015) <br> Insignificant (Kristoufek, 2015) |
| Hash rate | Positive (Kristoufek, 2015) <br> Positive (equilibrium model, Pagnotta, 2020) <br> Insignificant (Bouoiyour et al., 2015; Kristoufek, 2013) |
| Number of bitcoins | Insignificant (Kristoufek, 2015) <br> Negative (Ciaian et al., 2016; Goczek & Skliarov, 2019) |
| Unique bitcoin transactions per day | Insignificant (Ciaian et al., 2016) |
| Number of unique bitcoin addresses used each day | Insignificant (Ciaian et al., 2016) |
| Bitcoin velocity | Insignificant (Bouoiyour et al., 2015; Ciaian et al., 2016; Kristoufek, 2013) |
| Halving effect | Positive (Meynkhard, 2019; Fantazzini & Kolodin, 2020) <br> Indeterminant direction, equilibrium model, Pagnotta, 2020) |
| Cost of mining | Positive (Meynkhard, 2019; Hayes, 2019) |
| Transaction Network Edges | Positive (Kurbucz, 2019) |

## Sentiment and Autoregressive Effects on Bitcoin's Price

A significant amount of research has attributed the extremely high volatility of bitcoin's price to feedback mechanisms that reinforce the direction of price movements. As the bitcoin's price rises, social media and traditional news outlets broadcast stories about its upward trend, increasing interest, awareness, and fears of missing out (FOMO) on the potential returns of investing in the currency. When prices start to fall, stories about bitcoin's intrinsic lack of value and absence of backing can create near-panic selloffs. Autoregressive models and models tracking sentiment capture these feedback mechanisms very well. Table 3 references several representative articles that focus on or capture the psychological feedback mechanisms in the bitcoin market. For autoregressive integrative moving average (ARIMA) studies, we indicate whether the models are useful in predicting bitcoin's future value.

April 27, 2022

| Table 3 | |
|---|---|
| **Sentiment and Autoregressive Sources of Bitcoin Price Changes: US$ per Bitcoin** | |
| **Independent Variable** | **Results** |
| Investor interest/attractiveness | Positive, but price may lead interest (Bouoiyour et al., 2015; Bouoiyour & Selmi, 2015; Ciaian et al., 2016; Glaser et al., 2014; Kristoufek, 2013; Kristoufek, 2015); Yermack, 2013 |
| Tweet volume<br>Twitter sentiment<br>Social media sentiment and word of mouth (Reddit posts and new subscribers, Google search volume, Wikipedia views, Facebook shares)<br>Tweets; forum posts | Positive (Abraham, 2018)<br>Positive (Garcia et al., 2014)<br>Positive (Pant et al., 2018; Garcia et al., 2014;  Garcia et al., 2015; Phillips & Gorse, 2017)<br>Insignificant (Verma & Sharma, 2020)<br>Positive  (Mai et al., 2018; Kim et al., 2016; Phillips & Gorse, 2017; Bouoiyour et al., 2015; Bouoiyour & Selmi, 2015; Ciaian et al., 2016; Kristoufek, 2013, Pant et al., 2018; Yermack, 2013) |
| Autoregressive models | Useful (Azari, 2019; Garcia et al., 2014; Chevapatrakul & Mascia, 2019) |

Despite the many models that have been used to investigate bitcoin's price movements, there is little agreement as to what factors are most important and, in some cases, whether certain factors have positive or negative impacts on bitcoin's price. Attempts to resolve these differences have included separating bitcoin's history into two or more periods (Ciaian, 2016; Li & Wang, 2017), distinguishing between short-term and long-term impacts (Bouoiyour et al., 2015; Ciaian et al., 2016; Kristoufek, 2015; Li & Wang, 2017), and considering nonlinear formulations (Balcilar et al., 2017).

Relatively recently, machine-learning models, such as Random Forest, XGBoost, Quadratic Discriminant Analysis, Support Vector Machine and Long Short-term Memory, Recurrent Neural Network, and Particle Swarm Optimization, have been employed to predict the prices of cryptocurrencies, in general, and bitcoin, in particular. While the short-term results seem to be more robust than statistical models and analyses, machine-learning models for predicting long-term bitcoin exchange rates are still rather inaccurate (Chen et al., 2020, Indera et al., 2017, Rathan et al., 2019, and Velankar et al., 2018). Econometric work has also been done to determine if bitcoin is a hedge, diversifier, or safe-haven currency, but these analyses have focused more on correlation than causation (Bouri et al., 2017).

## Bitcoin Price Manipulation

The Bank for International Settlements (2019) reports that the *daily* turnover of fiat currencies exceeded $6.6 trillion. Nevertheless, there is clear evidence that manipulation of some fiat currencies has been widespread, driven in large part by the actions of governments and their central banks (Gagnon, 2012; Bergsten, 2013) – particularly those trying to fix their exchange rates to a dominant foreign currency. Despite the depth of these markets, speculative attacks on fiat currencies have also occurred (Buiter, 1987; Grilli, 1986), and investors have learned how to profit from currency manipulation (Liss, 2007).

Any currency that lacks liquidity is subject to manipulation, which is why cryptocurrencies have come under particular scrutiny. Manipulating bitcoin's price should require fewer resources than a dominant fiat currency because daily turnover is relatively low, and no central bank exists to protect it. Peterson (2021) concludes, with confidence greater than or equal to 95%, that illicit activity manipulated bitcoin's price in 2013, 2018, and 2019. Hu et al. (2020) analyzed intraday orders and trades and found significant evidence of bitcoin price manipulation during this cryptocurrency's price bubble in late 2017. Griffin and Shams (2020) claim that backers of the Tether stablecoin acted to manipulate bitcoin's price by issuing Tether without adequate backing, thereby inflating bitcoin's price. Wei (2018) finds, to the contrary, that Tether issuances had no impact on bitcoin returns.

## Supply and Demand Analysis

In competitive markets, prices are determined by the flow forces of supply and demand (Cecchetti and Schoenholtz, 2020; Marthinsen, 2020). The equilibrium exchange rate is at the intersection of a downward-sloping demand and upward-sloping supply. Like any freely fluctuating currency, bitcoin's price is measured by the cost to purchase it in fiat currencies, altcoins, or precious metals, and it is determined by the *flow of* supply and demand forces per period. Participants in the dollar-bitcoin market are, potentially, identical to those in the dollar-euro market, but their participation levels and scale of involvement are markedly different. The vast majority of bitcoin buyers are investors and speculators, who are incentivized mainly by *changes in* returns and *fluctuations* in expected risks and returns. Currently, the demand for bitcoins (by dollar holders) is very price inelastic because:

- Very few products are valued in bitcoins. Therefore, dollar holders have little or no reason to convert dollars to bitcoins to buy bitcoin-denominated goods or services. Those that convert dollars to bitcoins to buy goods and services do so to pay individuals and businesses that will accept them;

- A paltry volume of dollars is converted into bitcoins for bitcoin-denominated aid or gifts;

- Individuals may convert dollars to bitcoins for investment and speculative reasons, but typically, for financial investments, they do so based on *expected changes* in bitcoin's price;

- No government (yet) is a substantial source of bitcoin demand,[3] and

- No central bank (yet) fixes its exchange rate relative to bitcoin or is interested in stabilizing bitcoin's price (i.e., exchange rate in terms of a fiat currency or altcoin).

The significant sources of bitcoin's supply are distinctively different from its demand and significantly different from the typical supply in non-bitcoin-related foreign exchange markets. In contrast to demand, there are many reasons why bitcoin owners might want to supply bitcoins to purchase fiat currencies, such as dollars, euros, and yuan. For example in the dollar-bitcoin market,

- Anyone paid in bitcoins or receiving them in exchange might supply them to purchase dollars and afterward purchase dollar-denominated goods and services. Furthermore, fiat currencies, such as the dollar, are trusted, relatively stable, and have vast financial networks.

- Some companies, albeit few, compensate their employees in bitcoins, and a growing number of companies, restaurants, bars, and shops accept bitcoins. For risk management purposes, if most of these individuals and establishments convert their bitcoins into fiat currencies, such as dollars, to pay their bills, this supply source could be significant during periods when bitcoin's price is expected to fall.

---

[3] In September 2021, El Salvador adopted bitcoin as legal tender, but its total holdings amounted to fewer than BTC2,000. El Salvador's "official currency" is the U.S. dollar.

* Bitcoin holders who wish to invest in dollar-related financial investments could also be significant sources of bitcoin supply – especially those wishing to diversify their portfolios, mitigate risks, and cash out of bitcoin-denominated investments that have appreciated or are expected to depreciate.

* Individuals and businesses that own bitcoins but wish to make transfers denominated in dollars are potential (albeit likely small) sources of supply to the dollar-bitcoin market.

* Bitcoin has no central bank seeking to stabilize its value. The number of coins mined increases at a predetermined rate, which is currently 6.25 bitcoins per 10 minutes. Miners, who earn the newly created bitcoins, have two main options: holding their freshly acquired bitcoins or converting them into a fiat or altcoin currency. If they decide to hold them, the newly created bitcoins have no direct effect on bitcoins' supply to the dollar-bitcoin exchange market. If they convert the bitcoins to a non-dollar currency or altcoin, the impact on bitcoins' dollar price would be indirect through triangular arbitrage.

Concerning bitcoin's fiat price, such as dollars per bitcoin or euros per bitcoin, the supply should be upward sloping and demand downward sloping, thereby ensuring a stable equilibrium. At the same time, none of the dollar-bitcoin market participants (listed above) responds with alacrity to bitcoin price changes because so few goods and services are denominated in bitcoins, resulting in extreme price insensitivity. Bitcoin's price is far less critical for investors and speculators than *expected* price changes, which is an exogenous variable that shifts demand and supply rather than causing movements along these curves. Therefore, the extreme inelasticity of supply and demand, with respect to its price, guarantees that any shift in the supply of or demand for bitcoins will cause a disproportionate change in price relative to quantity traded per period. Therefore, until the bitcoin market's depth increases, its role as a reliable store of value will continue to be threatened.

Since its creation in 2009, fluctuations in bitcoin's dollar price have been caused almost entirely by demand movements. To the extent these demand-related movements can be predicted, bitcoin's future price might be forecastable using traditional econometric methods, but many exogenous shocks, such as earthquakes, hyperinflation, price bubbles, hacks, social sentiment, political regime changes, and exchange rate controls, are random. Nearly 80% of the bitcoins created, to date, are held by their owners, with little or no intention to sell, which is to say the impact of demand is not on the quantity of bitcoins created but on the quantity per period offered to the foreign exchange market. [4]

## Relationship between Mining Costs and Bitcoin's Price

The dollar-bitcoin market has two characteristics of a purely competitive market, a homogeneous asset and an absence of entry or exit barriers. Anyone wishing to buy or sell bitcoins can do so via exchanges or directly using the bitcoin protocol. Whether the bitcoin market has the final two characteristics of a

---

[4] Mark DeCambre, Who owns bitcoin? Roughly 80% are held by long-term investors: report, February 11, 2021, https://www.marketwatch.com/story/who-owns-bitcoin-roughly-80-are-held-by-long-term-investors-report-11612998740 (Accessed February 4, 2022).

April 27, 2022

purely competitive market (i.e., many buyers and sellers and perfect information) is debatable. The market is relatively shallow, which means a large-volume buyer or seller could significantly move the market price. Moreover, market information is not perfect, which is why, in the early days, arbitraging price discrepancies between exchanges had been possible. Recently, better information flows, caused by bitcoin's increased interest, have reduced or eliminated this ability.

Miners in the dollar-bitcoin market are price-takers and not price-makers. As a result, they face perfectly elastic demand curves, which means their average and marginal revenues are equal. If the average miner earns excess profits, then new miners should enter the market and reduce or eliminate these excess returns (Kristofek, 2020).

## How Excess Profits for Miners Affects Bitcoin's Price

On January 28, 2022, the total estimated bitcoin hash rate was approximately 204 exahashes per second, which equals 19,584 exahashes per bitcoin mined.[5] Bitcoin's price on that day was about $37,150, and miners were compensated at a rate of 6.25 bitcoins per 10-minute period. Given these conditions, a miner who controlled just 1% of the total bitcoin hashing power, such as the 58COIN Pool, would expect to earn daily revenues equal to $334,350.[6] With these expected revenues per day, our 1%-miner would pay his/her fixed and variable business expenses.

Variable expenses include costs, such as electricity and personnel, while fixed expenses include computers, new and faster processing chips, and office space (Kristoufek, 2020). A profit-maximizing miner should produce to the point where the marginal cost of production equals the marginal revenue. If excess profits are earned, new miners should enter, and existing miners should be incentivized to change their hashing capacities, depending on whether they are on the upward-sloping or downward-sloping portion of their long-term average total costs curves.

For two primary reasons, the entry of miners into or exit from the bitcoin foreign exchange market does not influence market supply and, therefore, does not affect bitcoin's price. First, bitcoin's total quantity increases at a predetermined rate, set by bitcoin's mathematical algorithm, rather than the number of miners. Currently, bitcoin's growth rate is BTC6.25 per 10-minute interval. If changes in miners' efficiency cause this growth rate to accelerate above or decelerate below the BTC6.25 per 10-minute pace, bitcoin's algorithm automatically adjusts the difficulty level to reset the target to BTC6.25 per 10-minute interval. The second reason bitcoin's fixed money creation has little or no effect on its fiat-currency price is because miners must offer the newly created bitcoins to the foreign exchange market for them to affect its price. If miners hold the newly created bitcoins, there is no direct effect on their fiat-currency price.

---

[5] An exahash is 1,000,000,000,000,000,000 (i.e., one quintillion) hashes. (204 exahashes/second) × (60 seconds/minute) × 10 minutes/block) ÷ (6.25 bitcoins/block) = 19,584 exahashes/bitcoin.

[6] This miner has a 1% chance of earning 6.25 bitcoins every 10 minutes and a 99% chance of earning nothing. Therefore, the expected revenue per 10-minute-interval equals [(1% × BTC6.25 × $37,150/BTC)] + (99% × BTC 0 × $20,000/BTC)] = $2,321.88 per 10-minute interval. Because a 24-hour day can be separated into 144 ten-minute segments, the expected daily return equals 144 x $2,321.88 = $334,350.

Consider the effect that market entry has on a miner who controls 1% of the total mining capacity. - Three major scenarios are possible.

- **Case #1**: Suppose that, despite increased competition, our 1% miner defends its market share by investing in greater hashing power. Expected revenues would remain the same because at bitcoin's market price and our miner's expected chances of success remain the same, but mining costs would rise, causing profits to fall.

- **Case #2**: Suppose our 1%-miner did not invest in new hashing power, thereby incurring no further costs. Under these conditions, its hashing share would fall as new miners entered and existing miners expanded capacity, causing expected revenues and profits to fall.

- **Case #3:** Finally, suppose our miner invested in greater hashing power, which increased its market share. Expected revenues would rise with the increased probability of success, but the effect on profits would be uncertain, depending on whether per unit costs rose or fell and whether revenues rose more or less than costs. If our miner could earn excess profits, then a new wave of miners would enter the market, and existing miners would be incentivized to expand capacity, thereby driving long-term excess gains to zero.

The miner's decision to adjust its scale of production would depend on the expected change in long-term average total costs. If they are expected to fall, the optimal strategy would be to increase mining activities. If the reverse were true, the miner would reduce its scale. Any changes in capacity that increased excess profits would encourage more competition, which would reduce earnings in the long run.

Using the dollar as the reference fiat currency, the main take-away points from these three cases are:

(1) Bitcoin's mathematical algorithm determines the rate at which the overall supply increases. The entry and exit of miners do not affect the rate at which new bitcoins are supplied to the dollar-bitcoin market;

(2) Only if newly created bitcoins are sold for dollars would they affect the dollar per bitcoin exchange rate;

(3) Improvements in technology, such as a new ASIC chip that doubles a miner's hashing power, do not affect the rate at which bitcoins are created because the bitcoin algorithm automatically adjusts the level of mining difficulty, thereby ensuring that no more than the predefined amount of bitcoins are created each period, and

(4) Due to these forces, any excess mining profits should be driven to zero in the long run by (1) the entry of new miners and changes in existing miners' hashing power, which reduces the expected chances of success and (2) rising costs, fueled by the need to maintain or expand hashing market share. Similarly, if operating losses occur, miners would exit the market, thereby increasing survivors' chances of success and reducing the demand for costly technological developments that increase hashing power.

(5) For exogenous shocks to the dollar-bitcoin market, the long-term effects on excess profits should be the same. For example, if energy costs rise, marginally profitable miners would drop out, increasing survivors' market share. As a result, their expected revenues would increase simultaneously as their energy costs rose, thereby driving excess profits to zero.

April 27, 2022

Appendix A formalizes this qualitative analysis mathematically. It shows that the average variable cost of mining a bitcoin in a given period depends on the price and excess profits per unit in the preceding period. As the price rises increasing excess profits, the cost of mining one bitcoin will rise in the following period assuming electricity costs remain stable. Similarly, as the price falls, decreasing excess profits, the cost of mining one bitcoin in the subsequent period will fall.

## Empirical Evidence

Recent empirical studies support our conclusion that cost changes have no significant impact on bitcoin price and follow price changes rather than the reverse. For example, Kristoufek (2020, p. 7), using a cointegration and vector error correction model for data from January 2014 to July 2018, concludes, "bitcoin price drives the mining costs and not (or only weakly) the other way around." Fantazzini & Kolodin (2020, p. 17), whose data run from August 1, 2015 to February 29, 2020, conclude, "there was neither evidence of Granger-causality nor cointegration in the first examined sample [from August 1, 2016 to December 4, 2017], …, whereas there was evidence of unidirectional Granger-causality and cointegration in the second sample [from December 11, 2017 to February 24, 2020] …, going from the bitcoin price to the hashrate (or to the [cost of production models]) but not vice versa." Kjærland, et al, whose data run from January 1, 2013 to February 20, 2018, find "… the technological factor Hashrate should not be included in modeling price dynamics or fundamental values since it does not affect Bitcoin supply." "[W]e believe that the causality between Bitcoin and Hashrate is such that it is the Bitcoin price that drives Hashrate, not the other way around. This outcome is consistent with economic theory since an increase in price will naturally result in the increased profitability of mining." (Kjærland, et al., 2018).

Our analysis does not predict bitcoin's price but rather uses economic theory to explain Kristoufek's (2020), Fantazzini & Kolodin's (2020), and Kjærland, et al.'s (2018) results. In this section, we clarify the logic behind why causation should run from changes in bitcoin's price to changes in the cost of producing bitcoins (i.e., hash rate) and not vice versa. Using data from January 3, 2017 to January 28, 2022, we show how bitcoin's hash rate adjusted to its exogenously changing price. This interval is divided into four periods, each of which captures a rise and fall in the USD price of bitcoin. The natural break between the periods can be seen most clearly by graphing the log of price as shown in Exhibit 1. The vertical bars between periods represent the low points between extended price decreases and increases.

**Exhibit 1**
**Bitcoin's Log Price: January 1, 2017 to January 28, 2022**

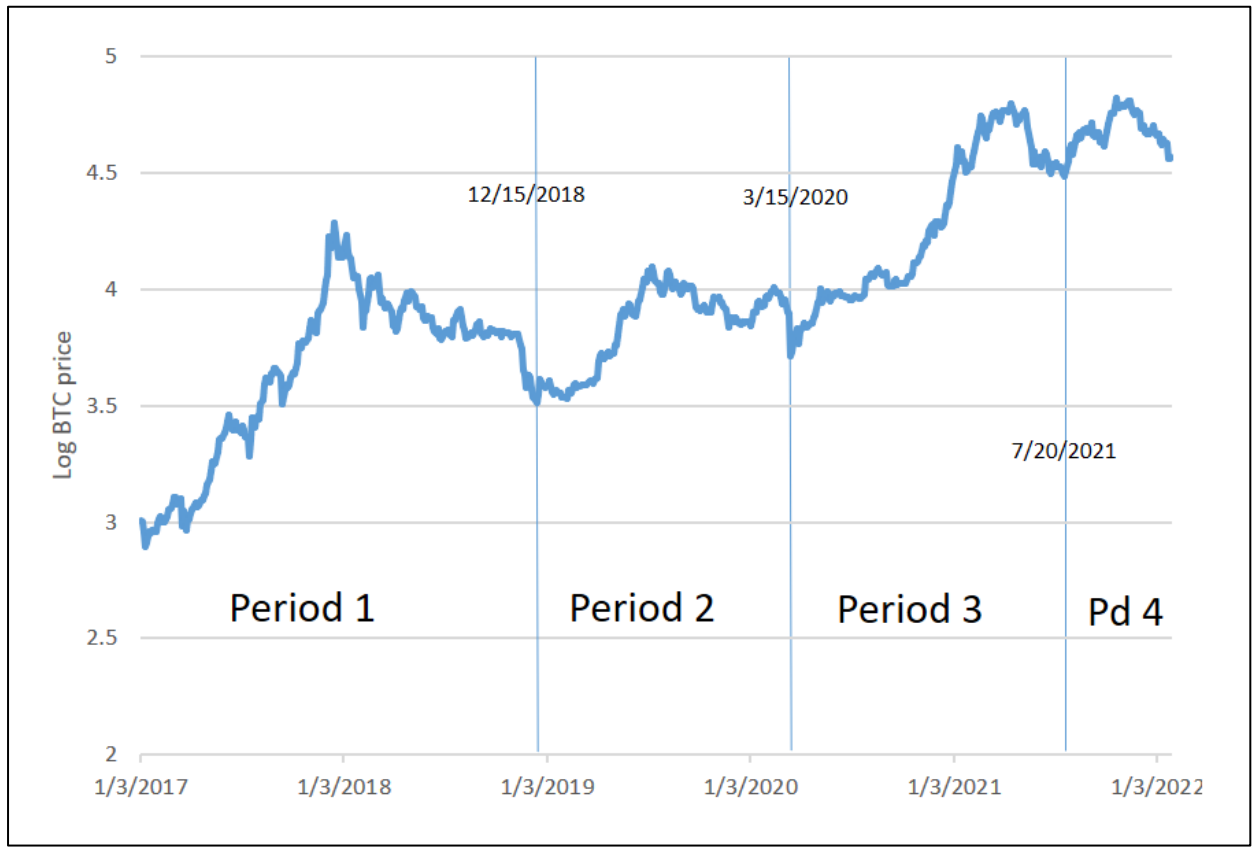

Exhibits 2 through 5 display, with appropriate scales for each period, the dollar price of a bitcoin versus the number of exahashes used to mine a bitcoin, which is a proxy for its cost. While both the price of a bitcoin and the number of exahashes to create one have increased over the years, a cursory examination of these exhibits demonstrates that there is little, if any, relationship between the two, and if anything, the hash rate follows the price dynamic rather than the reverse, as found by Kjærland, et al. (2018), Fantazzini & Kolodin (2020), and Kristoufek (2020). In Period 1, the hash rate was slow to rise as the price of bitcoin rose to a peak. [7] From December 17, 2017, to September 25, 2018, the price of bitcoin fell by 66% but its hashrate rose by 311%. This divergence reflects miners' reaction to the 2017 price bubble, which increased profits and incentivized others to enter the market and existing miners to boost their hash rates by more fully employing existing capacity and by purchasing and installing, with some lag, new, more efficient mining rigs. It took about nine months before bitcoin's price drop finally eliminated excess profits and forced miners to deactivate their least profitable rigs and reduce their hash rates in other ways. Period 2 shows the hash rate again rising slowly, as bitcoin's price peaked, and then continuing to rise despite bitcoin's price decline. Period 3 includes the bitcoin halving on May 11, 2020. Because the number of bitcoins awarded every ten minutes fell from 12.5 to 6.25, the number of hashes per bitcoin produced instantly doubled. This had practically no effect on the price of bitcoin, which remained steady for about five months even as the hash rate climbed. Eventually, the price of bitcoin bubbled, rising 490% from October 8, 2020, to April 15, 2021, but the hash rate failed to follow the bubble rising by only 28%. In Period 4, the hash rate shows a continuous rise whereas bitcoin's price rises and falls. The correlation between bitcoin's price and hash rate is less than 20% during this period.

---

[7] The introduction of a bitcoin futures market on December 10, 2017 did not have a significant impact on bitcoin's price during Bubble Period #1 but may have contributed to the decline in 2018. Empirical results are inconclusive. See Hattori and Ishida (2021) and Liu et al (2019), and Jalan et al. (2019). Fantazzini & Kolodin (2020) report that the regulated futures trades made the bitcoin market more mature and efficient.

**Exhibit 2**
**Bitcoin Price and Hash Rate during Period #1: January 3, 2017 to December 15, 2018**

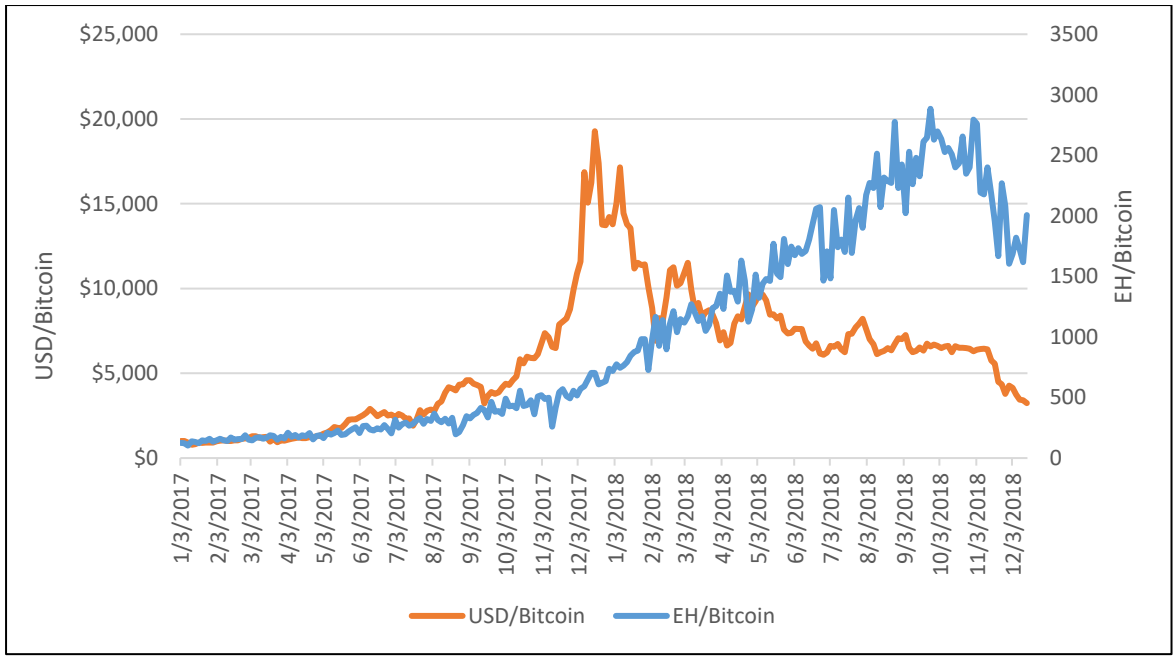

**Exhibit 3**
**Bitcoin Price and Hash Rate during Period #2: December 15, 2018 to March 15, 2020**

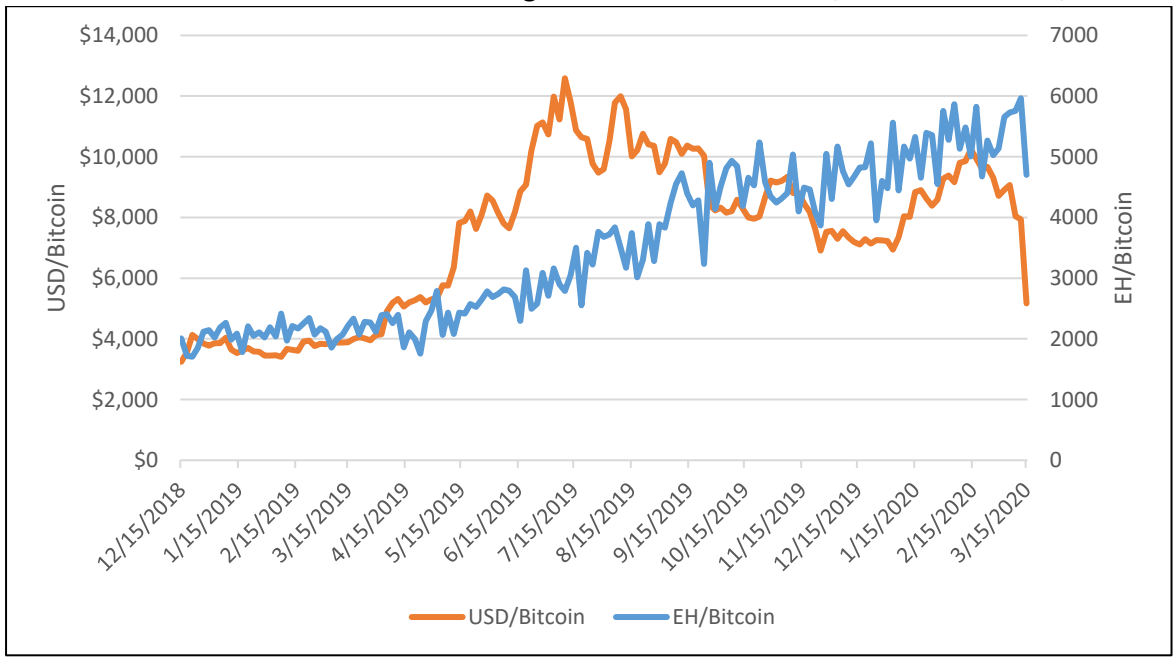

**Exhibit 4**
**Bitcoin Price and Hash Rate during Period #3: March 15, 2020 to July 20, 2021**

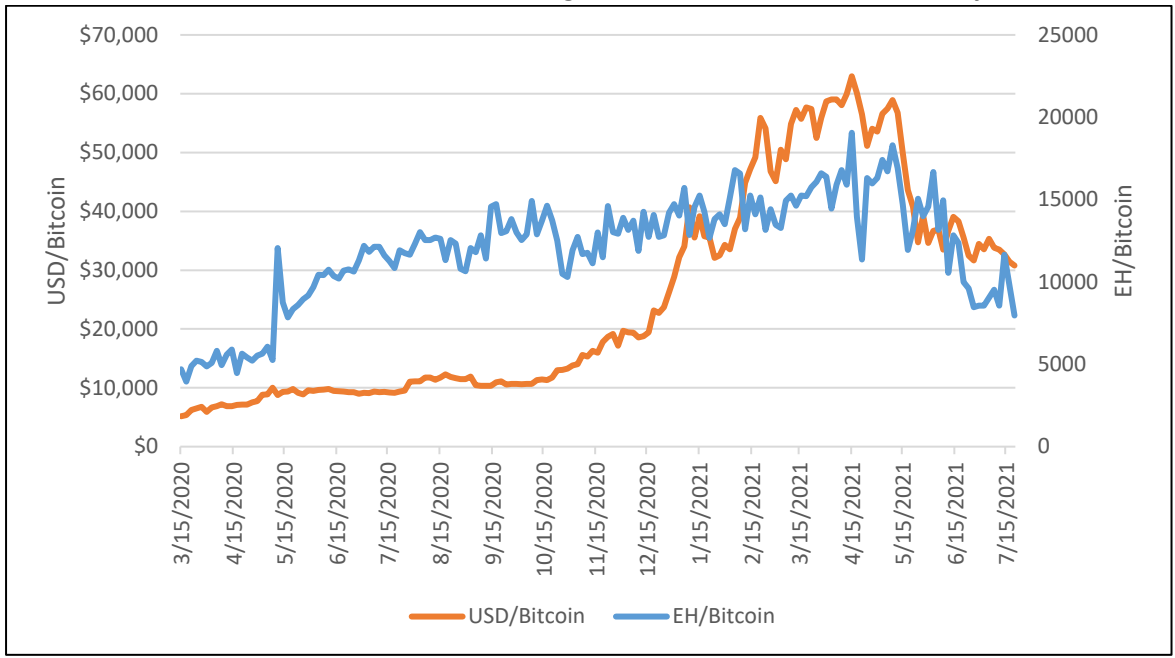

**Exhibit 5**
**Bitcoin Price and Hash Rate during Period #4: July 20, 2021 to January 28, 2022**

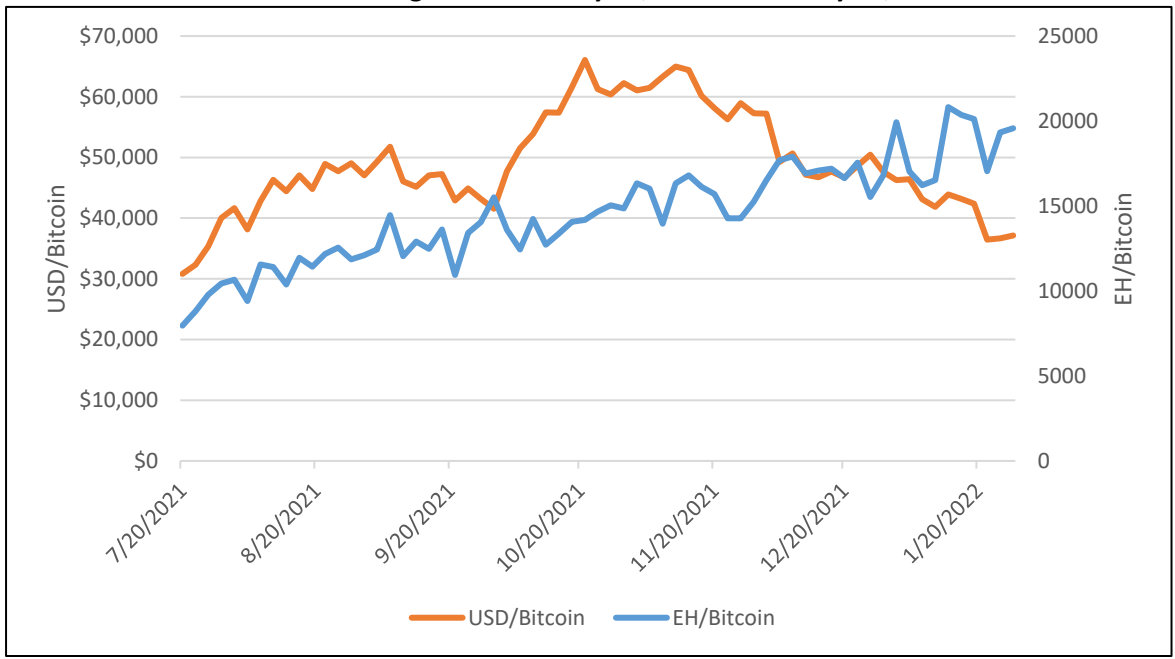

# Conclusion

Bitcoin has an independent "value" but no way to measure this value's change because there are no official or unofficial bitcoin price indices. Furthermore, its insignificant use as a unit of account, medium

April 27, 2022

of exchange, and store of value means that bitcoin fails to satisfy the major functions that "money" is supposed to fulfill. Therefore, care must be taken when couching analyses in the QTM or PPP context.

In contrast to its *value,* the *price* of a bitcoin can be measured easily and analyzed in the same way as foreign exchange rates. At its current level of development, bitcoin's supply and demand are highly inelastic. As a result, any significant exogenous shock to supply or demand, ceteris paribus, will cause volatile movements in price, thereby reducing bitcoin's chances of acting as a meaningful and reliable store of value.

Finally, bitcoin's production costs (i.e., mining cost) have no strong, direct effect on the bitcoin foreign exchange markets' supply and demand forces. In a curious turn of events, the cost of mining a bitcoin has virtually no impact on its price. Instead, causation is reversed, with increases (decreases) in bitcoin's price increasing (decreasing) miners' costs.[8] This result is consistent with Fantazinni and Kolodin (2020) and Kristoufek (2020), whose econometric studies find that "in the long run, mining costs strongly react to the increase in bitcoin's price and eventually catch up."

## Acknowledgements

We wish to thank the anonymous reviewers for their helpful and constructive suggestions.

---

[8] The focus of this paper is on why econometric studies, to date, have confused the cause-and-effect relationship between bitcoin's price and cost. In doing so, we considered the 2009-to-2021 period during which miner bitcoin rewards far exceeded transactions fees. We anticipate this will continue at least until 2028, when the bitcoin reward will be BTC 1.5625, which at current prices is worth more than $90,000 and far in excess of transactions fees. When new bitcoin creation stops or becomes sufficiently low, the cause and effect relationship we point out will end.

April 27, 2022

# Appendix A

Period $\quad$ = 10 minutes

$EX^0 \quad$ = Excess profits in Period 0 = Total Revenue$^0$ − Total Cost$^0$ = $TR^0 - TVC^0$

$n^1 \quad$ = Number of new hashes in Period 1 per dollar of excess profits in Period 0

$P^0 \quad$ = Price of bitcoin in Period 0

$f^0 \quad$ = Fees paid by transactions in Period 0

$TR^0 \quad$ = Total revenue in Period 0 = $6.25 \times P^0 + f^0$

$EL^i \quad$ = Electric cost in Period $i$ per hash unit

$\sum h_i^0 \quad$ = Sum of hashes in Period 0 by all miners

$TVC^0 \quad$ = Total Variable Cost in Period 0

$\qquad$ = Total number of hashes in Period 0 $\times$ Electric cost per hash in Period 0

$\qquad = \sum h_i^0 \times EL^0$

$\sum h_i^1 \quad$ = Sum of hashes in Period 1 by all miners

$\qquad$ = Total hashes in Period 0 plus $n^1$ times excess profits in Period 0.

$\qquad = \sum h_i^0 + (n^1 \times EX^0) = \sum h_i^0 + [n^1 \times (TR^0 - TVC^0)]$

$\qquad = \sum h_i^0 + n^1 \times [(6.25 \times P^0 + f^0) - (\sum h_i^0 \times EL^0)]$

$TVC^1 \quad$ = Total Variable Cost in Period 1

$\qquad$ = Total number of hashes in Period 1 $\times$ Electric cost per hash in Period 1

$\qquad = (\sum h_i^1 \times EL^1)$

$\qquad = [\sum h_i^0 + n^1 \times [(6.25 \times P^0 + f^0) - (\sum h_i^0 \times EL^0)] \times EL^1$